\documentstyle[12pt]{article}

\begin{document}
{\sf \begin{center} \noindent {\Large \bf Kerr-Schild Riemannian acoustic black holes in dynamo plasma laboratory}\\[3mm]

by \\[0.3cm]

{\sl L.C. Garcia de Andrade}\\

\vspace{0.5cm} Departamento de F\'{\i}sica
Te\'orica -- IF -- Universidade do Estado do Rio de Janeiro-UERJ\\[-3mm]
Rua S\~ao Francisco Xavier, 524\\[-3mm]
Cep 20550-003, Maracan\~a, Rio de Janeiro, RJ, Brasil\\[-3mm]
Electronic mail address: garcia@dft.if.uerj.br\\[-3mm]
\vspace{2cm} {\bf Abstract}
\end{center}
\paragraph*{}
Since Alfven, dynamo and sound waves and the existence of general
relativistic black holes are well stablished in plasma physics, this
provides enough motivation to investigate the presence of acoustic
black-hole effective metric of analogue Einstein's gravity in dynamo
flows. From nonlinear dynamo equations, one obtains a
non-homogeneous wave equation where it is shown that the
non-homogeneous factor is proportional to time evolution of the
compressibility factor. In the Navier-Stokes case for a finite
Reynolds number the acoustic black-holes also exists on the
stretching plasma flows. In the magnetostatic case the dynamo is
marginal. A coupled nonlinear plasma flow solution is found for the
dynamo equation where the effective black hole solution of the
scalar effective equation yields an imaginary part of the growth of
magnetic field. Therefore though the real part of the growth rate of
the magnetic field is negative or null, since there is a temporal
oscillation in magnetic field ,the solution represents a slow
dynamo. Thus acoustic black holes are shown to definitely contribute
to dynamo action of the effective plasma spacetime. It is suggested
that a fast dynamo effective spacetime may also contain an acoustic
black hole. I the case of planar waves the effective metric can be
cast in Kerr-Schild spacetime form. The Killing symmetries are
explicitly given in this metric and the growth of dynamo waves. A
further example of effective spacetime is given in collision
plasmas.{\bf PACS
numbers:\hfill\parbox[t]{13.5cm}{02.40.Hw:differential geometries.
91.25.Cw-dynamo theories.}}}

\newpage
\newpage
 \section{Introduction}
 Several attempts have very recently been made in building an artificial black hole in laboratory. In particular optical black holes \cite{1} have been
 obtained in optical fibers. Acoustic black holes are in general obtained as effective spacetime metric \cite{2} on a fluid media or flow. As in optical black holes
 a physical medium able to support some kind of waves are used as motivations to derive the acoustic or optical effective black holes. Since plasma medium is
 able to support this kind of waves along with Alfven waves and dynamo waves, it seems worth enough to investigate in this paper the effective metric of acoustic
 black holes in plasmas. Stretching of the
 magnetic field is in generally considered a fundamental ingredient for the dynamo \cite{3} existence. In this report however one shows that
 acoustic effective black holes analogue can be obtained on a non-dynamo dissipative plasma flow retangular slab. These so-acoustic black holes are effective
 analogue pseudo-Riemannian metrics given by the homogeneous wave equation from linearised Euler flows. Recently non-Riemannian vortex acoustic metrics
 \cite{4} in Navier-Stokes \cite{5} flows have been investigated by the
 author. In this paper one also show that the non-Riemannian geometry
 is not actually needed if one considers the pseudo-Riemannian acoustic spacetime
 non-homogeneous wave. Is exactly this non-homogeneous or
 non-Riemannian factor which yields
 the non-compressibility of the flow. Since is exactly this compressibility which is actually connected with stretching one may say that acoustic black holes
 would be obtained in stretching dissipative non-Riemannian flows while non-dissipative, either incompressible or compressible acoustic black holes could
 be naturally obtained. Magnetostatic fields are obtained in plasma dissipative flows. The stretching of the
 plasma flow is computes as well. Therefore Riemannian non-Riemannian manifolds can be used in this
 way to investigated effective metrics in plasma flows. One of the advantages of considering plasma slab in laboratory in the effective gravity
 is that in astrophysical settings the Reynolds number are very high as
 $Rm^{-1}\approx{{\eta}{\nabla}^{2}\approx{{\eta}{L^{-2}}}\approx{{\eta}{\times}10^{-20}}cm^{-2}}$
 for for example a solar loop scale length of $10^{10}cm$, diffusion effects neglected.  The paper is
 organized as
 follows: In section II a brief review on holonomic Frenet frame is presented where it is demonstrated that stretching of the flow is
 in generally associated with compressibility. In section
 III the dynamo equation is solved to yield an effective metric and a marginal dynamo is found. In section IV slow dynamos are obtained. Section V presents a further example
 of the effective plasma in the background of acoustic black holes. Section VI addresses the
 conclusions and discussions are
 given.
\newpage
\section{Stretching flows in Frenet frame} This section deals with a brief review of the Serret-Frenet
holonomic frame \cite{6} equations that are specially useful in the
investigation of fast dynamos in magnetohydrodynamics (MHD) with
magnetic diffusion. Here the Frenet frame is attached along the
magnetic flow streamlines which possesses Frenet torsion and
curvature \cite{6}, which completely determine topologically the
filaments, one needs some dynamical relations from vector analysis
and differential geometry of curves such as the Frenet frame
$(\textbf{t},\textbf{n},\textbf{b})$ equations
\begin{equation}
\textbf{t}'=\kappa\textbf{n} \label{1}
\end{equation}
\begin{equation}
\textbf{n}'=-\kappa\textbf{t}+ {\tau}\textbf{b} \label{2}
\end{equation}
\begin{equation}
\textbf{b}'=-{\tau}\textbf{n} \label{3}
\end{equation}
The holonomic dynamical relations from vector analysis and
differential geometry of curves by
$(\textbf{t},\textbf{n},\textbf{b})$ equations in terms of time
\begin{equation}
\dot{\textbf{t}}=[{\kappa}'\textbf{b}-{\kappa}{\tau}\textbf{n}]
\label{4}
\end{equation}
\begin{equation}
\dot{\textbf{n}}={\kappa}\tau\textbf{t} \label{5}
\end{equation}
\begin{equation}
\dot{\textbf{b}}=-{\kappa}' \textbf{t} \label{6}
\end{equation}
along with the flow derivative
\begin{equation}
\dot{\textbf{t}}={\partial}_{t}\textbf{t}+(\vec{v}.{\nabla})\textbf{t}
\label{7}
\end{equation}
From these equations and the generic flow
\begin{equation}
\dot{\textbf{X}}=v_{s}\textbf{t}+v_{n}\textbf{n}+v_{b}\textbf{b}
\label{8}
\end{equation}
one obtains
\begin{equation}
\frac{{\partial}l}{{\partial}t}=(-\kappa{v}_{n}+{v_{s}}')l\label{9}
\end{equation}
where l is given by
\begin{equation}
l:=(\textbf{X}'.\textbf{X}')^{\frac{1}{2}}\label{10}
\end{equation}
which shows that if $v_{s}$ is constant, which fulfills the
solenoidal incompressible flow
\begin{equation}
{\nabla}.\textbf{v}=0\label{11}
\end{equation}
since $v_{n}$ vanishes along the tube, one should have a
non-stretched flow. This is exactly the choice
$\textbf{v}=v_{0}\textbf{t}$, where $v_{0}=constant$ is the steady
flow one uses here. The solution
\begin{equation}
\textbf{B}={\nabla}{\phi}\label{12}
\end{equation}
shall be considered here. This definition of magnetic filaments is
shows from the solenoidal carachter of the magnetic field
\begin{equation}
{\nabla}.\textbf{B}=0\label{13}
\end{equation}
\section{Acoustic black holes in marginal dynamo plasmas}
Let us consider the non-linear dynamo flow equations \cite{7}
\begin{equation}
[{\partial}_{t}-Rm^{-1}{\nabla}^{2}]\textbf{B}={\nabla}{\times}(\textbf{v}{\times}\textbf{B})
\label{14}
\end{equation}
and the coupled dynamo flow equation
\begin{equation}
[{\partial}_{t}-Re^{-1}{\nabla}^{2}]\textbf{v}=-{\nabla}p+\textbf{J}{\times}\textbf{B}
\label{15}
\end{equation}
where $Re^{-1}$ and $Rm^{-1}$ are respectively the fluid and
magnetic Reynolds numbers and ${\textbf{J}}$ is the magnetic
current. Since the Reynolds flow numbers are finite, the magnetic
field is not necessarily frozen in and the flow may possess
magnetostatic fields. In the magnetostatic case one has
\begin{equation}
\textbf{B}={\nabla}{\phi} \label{16}
\end{equation}
and to obtain the effective metric as usual one considers the
irrotational flow \cite{7}
\begin{equation}
\textbf{v}={\nabla}{\psi} \label{17}
\end{equation}
First the magnetostatic equation decouples both equations above
since the current $\textbf{J}={\nabla}{\times}\textbf{B}$ vanishes.
Along the other magnetohydrodynamic equations
\begin{equation}
{\partial}_{t}{\rho}+{\nabla}.({\rho}\textbf{v})=0 \label{18}
\end{equation}
and the barotropic equation of state
\begin{equation}
p=p({\rho})\label{19}
\end{equation}
where p is the pressure. Now by considering the pressure and scalar
fluctuations \cite{7}
\begin{equation}
p=p_{0}+{\epsilon}p_{1}\label{20}
\end{equation}
\begin{equation}
{\psi}={\psi}_{0}+{\epsilon}{\psi}_{1}\label{21}
\end{equation}
\begin{equation}
{\rho}={\rho}_{0}+{\epsilon}{\rho}_{1}\label{22}
\end{equation}
Substitution of these fluctuations into the evolution flow equations
\begin{equation}
{\partial}_{t}[{c^{-2}}_{sound}{\rho}_{0}({\partial}_{t}{\psi}_{1}+\textbf{v}_{0}.{\nabla}{\psi}_{1}-Re^{-1}{\nabla}^{2}{\psi}_{1})]=
{\nabla}.[{\rho}_{0}{\nabla}{\psi}_{1}-{c^{-2}}_{sound}{\rho}_{0}\textbf{v}_{0}({\partial}_{t}{\psi}_{1}+\textbf{v}_{0}.{\nabla}{\psi}_{1})]
\label{23}
\end{equation}
which yields the equation
\begin{equation}
\frac{1}{\sqrt{-g}}[{\partial}_{\mu}(\sqrt{-g}g^{{\mu}{\nu}}g^{{\mu}{\nu}}{\partial}_{\nu}{\psi}_{1})]=Re^{-1}{\nabla}^{2}
{\partial}_{t}{\psi}_{1} \label{24}
\end{equation}
where ${\mu}=0,1,2,3$, are the spacetime of effective gravity. Here
one also uses the relation
$c_{sound}=\frac{{\partial}p}{{\partial}{\rho}}$. Now let us note
that the term on the RHS of this equation can be expressed as
\begin{equation}
{\nabla}^{2} {\partial}_{t}{\psi}_{1}= {\nabla}.
{\partial}_{t}{\nabla}{\psi}_{1}=
{\partial}_{t}({\nabla}.\textbf{v}_{1})\label{25}
\end{equation}
Now note that when the flow is incompressible
${\nabla}.\textbf{v}=0$ , or non-stretching this equation reduces to
\begin{equation}
{\partial}_{t}\frac{1}{\sqrt{-g}}[{\partial}_{\mu}(\sqrt{-g}g^{{\mu}{\nu}}g^{{\mu}{\nu}}{\partial}_{\nu}{\psi}_{1})]=0
\label{26}
\end{equation}
Therefore in the magnetostatic case the decoupling of effective
black hole in marginal dynamos can be easily obtained. These dynamo
like acoustic black holes can be called linear. In the next section
one presents the effective acoustic black holes are present in
non-linear slow dynamos.
\newpage
\section{Kerr-Schild effective slow dynamos spacetime}
In this section one shall consider the effect of the Riemannian
acoustic black hole metric may have on non-linear dynamos in
plasmas, where the back reaction of Lorentz magnetic force cannot be
neglected as in last section. The presence of plasma jets in
supermassive black holes , is more than enough motivation for the
investigation of the presence of analogue models in plasmas, as
discussed here. This time the dynamo flow coupling equations cannot
be decoupled anymore. Let us now consider the nonlinear dynamo
equation in the form
\begin{equation}
[{\partial}_{t}-Rm^{-1}{\nabla}^{2}]\textbf{B}=(\textbf{v}.{\nabla})\textbf{B}-(\textbf{B}.{\nabla})\textbf{v}
\label{27}
\end{equation}
Let us now show that the acoustic black hole metric given by
$g^{00}=-\frac{1}{{c^{2}}_{sound}{\rho}_{0}}$,
$g^{0j}=-\frac{1}{{c^{2}}_{sound}{\rho}_{0}}{v^{j}}_{0}$ and
$g^{ij}=\frac{1}{{c^{2}}_{sound}{\rho}_{0}}({c}_{sound}({\delta}^{ij}-{v_{0}}^{i}{v_{0}}^{j}))$
and the magnetic field perturbation
$\textbf{B}=\textbf{B}_{0}+{\epsilon}\textbf{B}_{1}$ where
${\epsilon}<<1$ implies that the coupled dynamo flow equation
becomes
\begin{equation}
[\textbf{Re}{\gamma}+i\textbf{Im}{\gamma}+Re^{-1}K^{2}]\textbf{B}_{0}=i({\nabla}{\psi}_{0}.{\nabla}(\textbf{K}.\textbf{x}))\textbf{B}_{0}
\label{28}
\end{equation}
where $\textbf{x}=x\textbf{i}+y\textbf{j}+z\textbf{k}$ is the vector
describing points in the retangular coordinates of the slab and
$\textbf{K}$ is the wave vector of the ansatz for the magnetic field
\begin{equation}
\textbf{B}_{0}=exp[{\gamma}t+i(\textbf{K}.\textbf{x})]\textbf{b}_{0}
\label{29}
\end{equation}
and from the equation $div\textbf{B}=0$ one obtains
$\textbf{B}_{0}.\textbf{K}={0}$ which in turn yields
\begin{equation}
({\partial}_{x}+{\partial}_{y})\textbf{v}_{0}=0 \label{30}
\end{equation}
where
\begin{equation}
\textbf{B}_{0}=exp[{\gamma}t+i(\textbf{K}.\textbf{x})]\textbf{b}_{0}
\label{31}
\end{equation}
Thus from the acoustic metric one obtains finally an expression for
the metric in terms of the dynamo growth rate as
\begin{equation}
\textbf{Im}{\gamma}=({\nabla}{\psi}_{0}.{\nabla}(\phi)) \label{32}
\end{equation}
\begin{equation}
\textbf{Re}{\gamma}=-Re^{-1}K^{2} \label{33}
\end{equation}
Note from (\ref{33}) that the dynamo is not at all fast since
$\textbf{Re}{\gamma}<0$. Actually since the imaginary part of the
magnetic field
\begin{equation}
\textbf{Im}\textbf{B}_{0}=\textbf{b}_{0}sin[\textbf{Im}{\gamma}t+{\phi}]
\label{34}
\end{equation}
Here ${\phi}=\textbf{K}.\textbf{x}$ is the magnetic phase. Thus from
Equation (\ref{32}) is given by
\begin{equation}
v^{0j}{\partial}_{j}{\phi}=v^{0j}{K}_{j}\label{35}
\end{equation}
the acoustic metric one obtains finally an expression for the metric
in terms of the dynamo growth rate as
\begin{equation}
g^{0j}={c}_{sound}{\rho}_{0}\frac{\textbf{Im}{\gamma}}{{K}^{2}}{K}^{j}
\label{36}
\end{equation}
where $K^{2}=K_{j}K^{j}$. The other acoustic black hole dynamo
metric are
\begin{equation}
g^{ij}=\frac{1}{{\rho}_{0}}{c^{-2}}_{sound}({c_{sound}}{\delta}^{ij}-[Im{\gamma}]^{2}\frac{{\partial}^{i}{\phi}{\partial}^{j}{\phi}}{{K}^{4}})
\label{37}
\end{equation}
Note that when one considers planar magnetic waves in plasmas or
dynamo waves where the wave vector $\textbf{K}$ is constant one
obtains the Kerr-Schild spacetime metric
\begin{equation}
g^{ij}=\frac{1}{{\rho}_{0}}{c^{-2}}_{sound}({\delta}^{ij}-[Im{\gamma}]^{2}\frac{{K}^{i}K^{j}}{{K}^{4}})
\label{38}
\end{equation}
This completes our task of showing that an artificial acoustic black
hole can be obtained in the effective plasma slow dynamo spacetime.
Metric (\ref{38}) resembles metrics in scalar gravity and is
definitly a spacetime carrying the growth of magnetic field
information from $Im{\gamma}$ term. Actually $K^{j}$ is the Killing
vector of the Kerr-Schild stationary metric symmetry so much useful
in general relativistic metrics. \newpage
\section{Acoustic black
holes in collision plasmas} In this section a further physical
example is given of the Einstein's effective gravity in plasmas
dissipative media. Let us then to consider the equation including
the dissipation effective and the electric field as
\begin{equation}
{\rho}[{\partial}_{t}{\textbf{v}}+({\textbf{v}}.{\nabla})\textbf{v}]=\pm{en}\textbf{E}-{\rho}{\nu}\textbf{v}-{\nabla}p
\label{39}
\end{equation}
here $\textbf{E}$ is the electric field and ${\nu}$ is the viscosity
coefficient. From the same reasoning of the last section one
has
\begin{equation}
{\partial}_{t}[{c^{-2}}_{sound}{\rho}_{0}({\partial}_{t}{\psi}_{1}+\textbf{v}_{0}.{\nabla}{\psi}_{1}-Re^{-1}{\nabla}^{2}{\psi}_{1})]
\pm{en}{{\phi}_{1}}-{\rho}{\nu}{\psi}_{1}-p_{1}=
{\nabla}.[{\rho}_{0}{\nabla}{\psi}_{1}-{c^{-2}}_{sound}{\rho}_{0}\textbf{v}_{0}({\partial}_{t}{\psi}_{1}+\textbf{v}_{0}.{\nabla}{\psi}_{1})]
\label{40}
\end{equation}
which yields the equation
\begin{equation}
\frac{1}{\sqrt{-g}}[{\partial}_{\mu}(\sqrt{-g}g^{{\mu}{\nu}}g^{{\mu}{\nu}}{\partial}_{\nu}{\psi}_{1})]=\pm{en}{{\phi}_{1}}-{\rho}{\nu}{\psi}_{1}\label{41}
\end{equation}
where ${\mu}=0,1,2,3$, are the spacetime of effective gravity. Here
one also uses the relation
$c_{sound}=\frac{{\partial}p}{{\partial}{\rho}}$. Now let us note
that by expressing ${\psi}_{1}$ as
\begin{equation}
{\psi}_{1}=e^{{\omega}t+i(k_{r}r)}\label{42}
\end{equation}
and by considering that the background effective (2+1)-D spacetime
Riemannian line element is given by Visser fluid \cite{8} metric
\begin{equation}
ds^{2}=-{c^{2}}_{sound}dt^{2}-(dr-\frac{A}{r}dt)^{2}+(rd{\theta}-\frac{B}{r}dt)^{2}
\label{43}
\end{equation}
one obtains the following dispersion relation
\begin{equation}
[{\rho}_{0}({\omega}-{\nu})+\frac{AK_{0}}{r}]{\psi}_{1}=\pm{en}{{\phi}_{1}}-p_{1}\label{44}
\end{equation}
From the base part of the fluctuation equation
\begin{equation}
\pm{en{\phi}_{0}}=[{\rho}_{0}({\omega}+{\nu})-\frac{A}{r}K_{r}]{\psi}_{0}
\label{45}
\end{equation}
Since by the equations one has
\begin{equation}
{\psi}_{0}=-Alnr \label{46}
\end{equation}
Substitution of this expression into (\ref{45}) allows us to
determine the electric potential of the collision plasma
\begin{equation}
\pm{en{\phi}_{0}}=-A[{\rho}_{0}({\omega}+{\nu})-\frac{A}{r}K_{r}]lnr
\label{47}
\end{equation}
This and other interesting physical properties of this plasma
effective spacetime may be considered elsewhere.
\newpage
\section{Conclusions} A further example of an artificial
Riemannian black hole in the Euclidean geometry of plasma dynamo
slabs in retangular coordinates, is added to inumerours examples of
artificial optical and electromagnetic effective spacetimes
\cite{3}. other geometries can be considered elsewhere which comes
from plasma physics as the construction of the artificial black
holes in toroidal magnetic geometry of tokamaks and stellarators.
The toric geometry has actually been considered by Garay \cite{8} in
the investigation of Bose-Einstein condensates. An interesting
example of an artificial magnetic metric seems to have given by
Titov \cite{9} where he considers the case of covariant formulation
of solar loops as well but certainly with different details and
motivation. Actually Titov effective metric is more similar to
Novello´s, in reference \cite{8}, dielectric nonlinear metric
$g^{{\mu}{\nu}}={\epsilon}{\eta}^{{\mu}{\nu}}-\frac{{\epsilon}'}{E}(E^{\mu}E^{\nu}-E^{2}v^{\mu}v^{\nu})$
where $E^{\nu}$ is the electric field and ${\epsilon}$ is electric
permittivity and ${\epsilon}'=\frac{d}{dE}{\epsilon}$. This metric
and Titov's possesses the Kerr-Schild form for rotating black holes.
To resume since general relativistic plasma physics \cite{10} in
black hole physics is well stablished along with the presence of
sound waves in plasmas , the presence of acoustic black holes is
naturally investigated in this paper.
\section{Acknowledgements} I thank financial supports from Universidade do Estado do Rio de
Janeiro (UERJ) and CNPq (Brazilian Ministry of Science and
Technology).

  \end{document}